\documentclass[twocolumn,showpacs,preprintnumbers,amsmath,amssymb]{revtex4}
\usepackage{hyperref}
\usepackage{graphicx}
\usepackage{amsmath}
\usepackage{titlesec}
\usepackage{feynmp}
\usepackage{float}
\usepackage{epstopdf}
\titleformat{\section}[runin]{\bfseries}{}{}{}
\begin{document}

\title{Localization and Fractionalization in a Chain of Rotating Atomic Gases}
\author{Jianshi Zhao}
\author{Louis Rene Jacome}
\author{Craig Price}
\author{Nathan Gemelke}
\address{The Pennsylvania State University \\ University Park, PA 16802, USA}

\begin{abstract}  We describe the phase diagram and thermodynamic properties of a chain of axially-tunnel-coupled fractional quantum Hall systems realized by rotating a series of optical dipole traps about their center. We demonstrate not only a experimentally feasible pathway to a state describable as a Mott-insulator of composite bosons, but also describe the nature of the coherent states at higher tunnel coupling strength, and identify a series of new superfluid phases with rich behavior.  The phase diagram directly reveals not only characteristic features of the few-body systems, including the effective mass of composite particle- and hole-like excitations and their interactions, but emergent properties of the chain also reveal a fundamental mapping between the adiabatic dynamics of two-dimensional systems governed by particle braiding and the hydrostatic response of the gas in the conducting phases.
\end{abstract}
\maketitle

Shortly after the discovery of fractional quantum Hall effects in electronic systems, it was realized certain thermodynamic ground states possess excitations with fractionalized mass and statistics. Despite intense efforts, direct observation of these features has proved elusive in experiments on electronic systems. Recently, it was predicted \cite{Cooper99,Cooper01,Wilkin2000,Regnault03,Chang05b} that rapidly rotating gases of ultracold Bosonic atoms exhibit states similar to fractional quantum Hall ground states with emergent topological and quantum order, and experiments have begun to probe gases in this regime \cite{Gemelke10}. The question then arises as to how one can create and probe excitations in these systems which may shed light on the structure of the fractional Hall ground states, or perform other experiments indicative of this novel type of order. In a recent letter\cite{Zhang14}, some of us proposed that hole-like excitations in a fractional Hall state could be created and probed by introducing impurity atoms of a second atomic species, and that the pair-correlations of two such excitations reveal fractionalized angular momentum as a result of their fractionalized statistics. Here we consider a second method of interrogating fractional Hall effects in cold gases, in which many such samples are created along a chain of lattice sites, and coupled together via tunneling. We find that such systems support novel insulating and superfluid states, and that an interplay of the conserved quantities, particle number and angular momentum, arises, affecting the character and dynamics of quantum and thermal fluctuations, and leading to novel transport properties. We calculate a mean-field phase diagram and derive effective field theory to describe this system.

We consider the thermodynamic ground state of the Hamiltonian $\hat{\mathcal{H}}=\hat{\mathcal{H}}_\Omega+\hat{\mathcal{H}}_{||}-\mu \hat{N}$, where $\hat{\mathcal{H}}_\Omega=(1-\Omega/\omega)\hat{L}+\eta \hat{V}+\hat{\mathcal{H}}_\epsilon$ is the on-site hamiltonian for gas harmonically trapped and rotating at the frequency $\Omega$, projected onto the lowest-landau-level (LLL), and $\hat{\mathcal{H}}_{||}=-t\sum_{m,i}(\hat{a}^\dagger_{i+1,m}\hat{a}_{i,m} + \text{c.c.})$ describes tunneling along the chain. Here, the $\hat{a}_{i,m}$ destroy a particle of angular momentum $m$ at site $i$, and obey bosonic commutation relations $[\hat{a}_{i,m},\hat{a}^\dagger_{i',m'}]=\delta_{i,i'}\delta_{m,m'}$, and all energy scales are measured relative to the harmonic trap energy $\hbar\omega$ in the plane of rotation. The chemical potential $\mu$ is introduced to allow variation of the total particle number $\hat{N}=\sum_{i,m} \hat{a}^\dagger_{i,m}\hat{a}_{i,m}$. The total angular momentum $\hat{L}=\sum_{i,m} m \hat{a}^\dagger_{i,m}\hat{a}_{i,m}$, and the interaction energy is assumed to be given by contact interactions, whose form is given in the supplementary material. Finally, we include the effect of a small rotating quadrupolar moment to the local trapping potential through $\hat{\mathcal{H}}_\epsilon=\sum_{m,i}\epsilon\,(\hat{a}^\dagger_{i,m+2}\hat{a}_{i,m}+\text{c.c.})$, such that the major and minor trap frequencies $\omega_{\pm}$ give $\epsilon=2(\omega_+-\omega_-)/(\omega_++\omega_-)$.

\begin{figure}[htp]
\includegraphics[width=2.5 in]{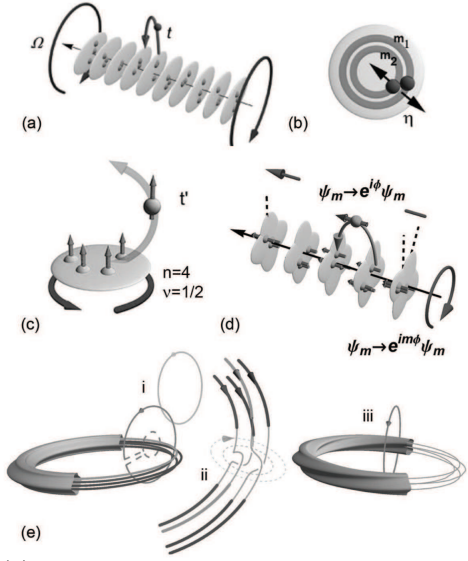}\vspace{-.25in}
\caption{(a) We consider a chain of rotating traps, through which atoms can tunnel at a rate $t$ along the rotation axis. At sufficiently high rotation rates $\Omega$, and with (b) sufficiently strong scattering due to repulsive interactions parameterized by $\eta$, atoms may occupy higher angular momentum eigenstates $m_i$ in the lowest Landau level, forming (c) strongly correlated fractional Hall states within in a two-dimensional well (filling factor $\nu=1/2$ state is illustrated for $n=4$ atoms). Such states are describable as composite particles bound to fluid vortices, and we show for moderate $t$ that tunneling between such states reduces to highly collaborative tunneling of composites at a renormalized rate $t'$. (d) Under appropriate conditions, a tunnel-coupled chain will form insulating states of well-defined atom number and FQH filling factor at each site, and superfluid states formed by local superpositions thereof. Such states possess novel transport properties, in which particle flow along the chain is described by motion of composite entities, which for a given axial flow induce a torsional strain in the background insulator. The linkage between flow and strain can be described in a chain with periodic boundary conditions (e) as a topology-preserving insertion (steps i-iii) of motional flux in the presence of intrinsic flux representing the insulating state.}\label{fig::fig1}
\end{figure}

\begin{figure}[htp]
\includegraphics[width=3.25 in]{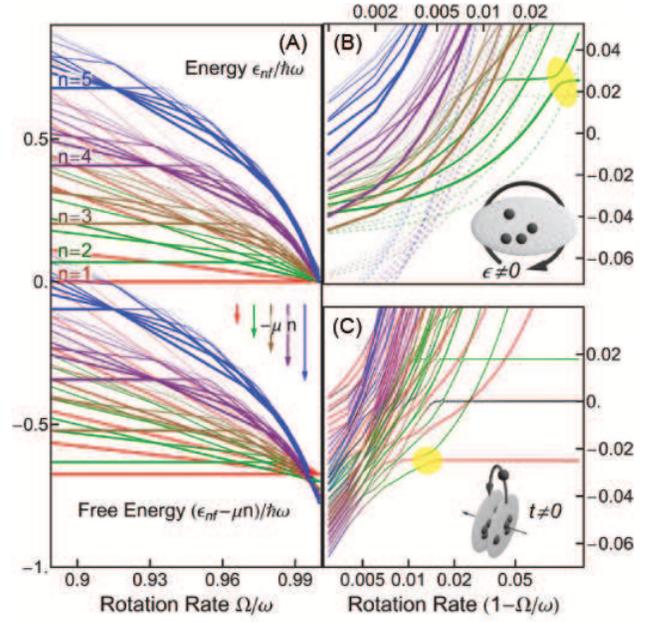}
\caption{The energy spectrum $\epsilon_{nf}$ for each of these $n$-body states (A) is shifted downward proportionate to the chemical potential $\mu$ to form the free energy in a grand-canonical ensemble. Experimental control over the many-body state can be exerted through (B) deformation of the rotating trap (strength $\epsilon$), which couples states of differing angular momentum (proportional to slope $d\epsilon_{nf}/d\Omega$) and introduces avoided level crossings (highlighted in yellow), or (C) by controlling the tunneling rate $t$, producing avoided crossings between states of differing particle number (expectation value of $n$ represented by color). Spectra shown here correspond to $\eta=0.0054$, $\mu=0.025\hbar\omega$, $\epsilon=0.002$ (in B), $t=0.004\hbar\omega$ (in C). Spectra in (A) and (B) were produced with DD, and (C) by minimization of the Gutzwiller form as described in the text.}\label{fig::fig2}
\end{figure}

The few-body eigenstates for the Hamiltonian $\hat{\mathcal{H}}_\Omega$ are calculated by direct diagonalization (DD), similar to the methods used in references \cite{Gemelke10,Paredes2001}. In figure \ref{fig::fig2}, we show the lowest eigen-energy of $\mathcal{H}_\Omega$ as a function of $\Omega$ for a number of atoms $n \leq 5$. As the rotation rate is increased, the kinetic energy penalty for single atoms to occupy higher angular momentum eigenstates is reduced, and interaction begins to mix the non-interacting eigenstates. As a result, at high $\Omega$, each occupancy enters into a series of progressively more strongly entangled ground states of higher angular momentum. Previously, we have identified the first and last states in this sequence as an $|L=n\rangle$ single-vortex, and $|L=n(n-1)\rangle$ $\frac{1}{2}$-Laughlin state, respectively, for all particle numbers (for $n=2$ these states are equivalent); other states have been tabulated extensively in previous literature, and experiments have now entered a regime in which individual few-body samples may be brought with reasonable fidelity into many of them. The $\frac{1}{2}$-Laughlin state is a generalization of the $\frac{1}{q}$-Laughlin state known from 2DEGs, whose wavefunction can be written in complex coordinates as $\phi_L({z_i})=\prod_{ij}(z_i-z_j)^2\exp(-\sum_i |z_i|^2/2)$. If the particle number in the $\frac{1}{2}$-Laughlin state is made indefinite through a coherent superposition of such states, it can be considered as a condensate of composite objects, in which $q$ quanta of vorticity (or equivalently a dynamical Chern-Simons gauge field as described in the supplementary material) are attached to each particle coordinate prior to condensation. In this letter, we interrogate whether such a state can be directly created in a chain of cold-atom FQH systems, and whether its dynamics are best described by the tunneling of composite bosons through the chain, or by the motion of bare particles.  We find that not only does the composite language form a good description, but that relations between emergent hydrostatic properties (such as compressibility and rotational moment and superfluid density and torsional stiffness) reflect a basic mapping of topology from two spatial plus one time (2+1) dimensions into a static three-dimensional version (3+0).

We form a mean-field model first by assuming an approximate Gutzwiller form for the many-body state of the system $|\Psi\rangle=\prod_j |\Psi\rangle_j$ with $|\Psi\rangle_j=\sum_{n,f} a_{n,f} |n,f\rangle_j$, where $a_{n,f}$ represents the amplitude of the $f^{th}$ lowest energy-eigenstate $|n,f\rangle_j$ of the $n$-atom solution to $\hat{\mathcal{H}}_\Omega$, localized on site $j$. Forming the energy $\langle \Psi |\hat{\mathcal{H}}| \Psi \rangle$, and minimizing its variation with $a_{n,f}$ subject to the constraint $\sum_{n,f} |a_{n,f}|^2=1$, one finds the condition $\sum_{n'f'} M_{\begin{smallmatrix}n\!n'\\f\!f'\end{smallmatrix}} a_{n'f'}\!\!=\!\!\mathcal{E}a_{nf}$, with

\begin{eqnarray}
M_{\begin{smallmatrix}n&\!n'\\f&\!f'\end{smallmatrix}}\!=\delta_{\begin{smallmatrix}n&\!n'\\f&\!f'\end{smallmatrix}}\![\epsilon_{n'f'}-\!\mu n']\!&\!-\!&\!2t\delta_{n'\!,n-1}\!\!\sum_m \psi_m\tau^{(m)*}_{n',f',f}+ \text{c.c.} \nonumber  
\label{eq::MFVGW}\end{eqnarray}

\begin{figure}\includegraphics[width=3.35 in]{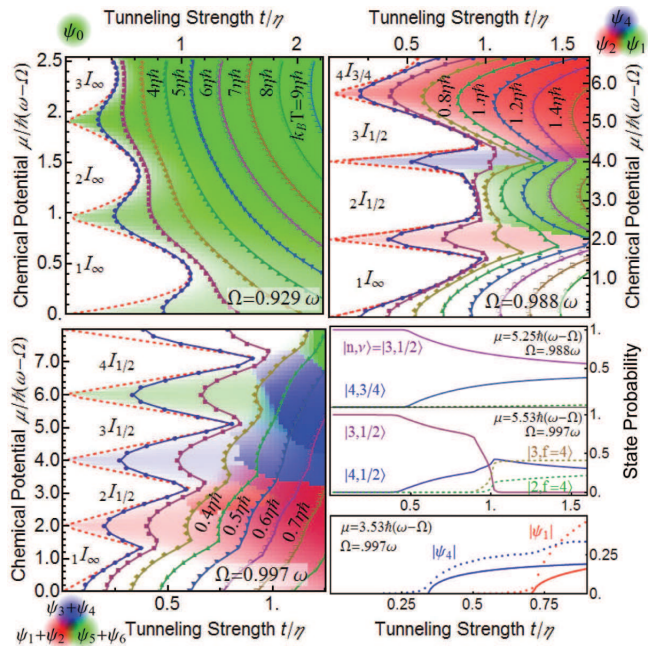}\caption{(A-C) Mean-field phase diagrams, showing components $\psi_m$ of the order parameter in color, and zero- and finite-temperature phase boundaries (joined symbols) at three rotation rates. The dashed red line shows the analytic result for the zero temperature phase boundary. (A, $\Omega=0.929\omega$) At low rotation rates, the phase-diagram reduces to that of the Bose-Hubbard model, exhibiting insulator-to-superfluid $_{n}I_{\infty}-SF$ transitions in the $m=0$ channel. (B, $\Omega=0.988\omega$) As the rotation rate is increased, high-occupancy insulating regions $_{n}I_\nu$ transition to higher angular momenta and stronger correlation due to the change in lowest-energy few-body form. (C, $\Omega=0.997\omega$) Near the centrifugal limit, insulators at all occupancies have transitioned into the $\nu=\frac{1}{2}$ (Laughlin) form, in which the on-site interaction energy is zero; at higher $t$, superfluidity develops as coherent superpositions of these states. When $t$ is sufficiently large, the Laughlin-superfluid is destroyed in favor of a more completely mean-field type state. (D-E)  Mean-field ground state probabilities $|a_{nf}|^2$ are shown as a function of tunneling strength $t$ for two different cuts from (B) and (C). They show two sudden changes, first at the insulator boundary, and a second jump as the FQH-forms are lost at higher $t$ - the latter can be roughly described by calculation of $u$ in an effective field theory, corresponding to saddle points in the free-energy along a second component of $\psi$ (points: numeric Gutzwiller ansatz, solid lines: EFT to fourth order in $\psi$.)}\label{fig::fig3}
\end{figure}

\noindent Here, the quantity $\mathcal{E}$ is a Lagrange multiplier, and $\epsilon_{nf}$ represents the energy of the state $|n,f\rangle_j$ due to $\hat{\mathcal{H}}_\Omega$. The nonlinearity is represented through the fields $\psi_m=\, _j\langle \Psi | \hat{a}_{j,m} | \Psi \rangle_j $, and the quantity $\tau^{(m)}_{n,f_1,f_2}\equiv \langle n, f_1 | \hat{a}_m | n+1, f_2 \rangle$ reflects the overlap of two few-body states when a single particle of angular momentum $m$ is removed, describing the effectively lower rate of tunneling between strongly correlated states on neighboring sites. A numeric solution of these equations can be performed by choosing an arbitrary set of coefficients $a_{nf}$, calculating the fields $\psi_m$, and solving for the lowest eigenvector of $M$, iterating until the fields converge \cite{Hazzard2010} - a representative sample of field values obtained this way are shown in figure \ref{fig::fig3}.

In the decoupled limit $t=0$, the many-body states are simultaneous eigenstates of the number operators $\hat{n}_j$ at each site $j$, and for $\epsilon=0$, are also eigenstates of angular momentum $\hat{L}_j$; and thus the ground-state wavefunction is a product state $\prod_j | n, L \rangle_j$ with $n$ and $L$ chosen to minimize the total energy $E=N_s (\epsilon_{n,L}-\mu n)$, reflecting a generalized Mott-insulator state, which we label as $_nI_\nu$ with filling factor $\nu=n(n-1)/2L$. For each particle number $n$, the ground-state crossing sequence consists of $n-1$ transitions, and thus below the chemical potential corresponding to $n_m$ filling, there are $n_m(n_m+1)/2$ such insulating states which can be reached through control of $\Omega$ and $\mu$. For sufficiently high rotation rates, the few-body ground state for any particle number $n$ is a $\nu=\frac 12$ ($\frac 12$-Laughlin) form, with zero interaction energy, and the total energy can be written as $\epsilon_{n,L=n(n-1)}=(1-\Omega/\omega)n(n-1)$. The boundaries between insulators $_nI_{1/2}$ therefore occur at the critical chemical potentials of $\mu_c=2n\hbar(\omega-\Omega)$. Adiabatic evolution or cooling of a sample into these or other insulator phases by dynamic control of $\Omega, \mu$ and $t$ provides an experimental method for producing large numbers of pure few-body FQH states of definite occupancy and filling factor.

One might naively guess that the transition at the insulator boundary signifies the onset of superfluidity of composite particles along the length of the chain, forming a locally coherent state. In fact, if one retains only the lowest few-body eigenstates just below the centrifugal limit, a mean-field calculation would support this picture somewhat trivially, as described in reference \cite{Hazzard2010}. However, retaining the full few-body spectrum for each particle number, the mean-field calculation presented in figure~\ref{fig::fig3}C does not support this picture beyond sufficiently high $t$, nor does it strictly support the picture of the superfluid phase as a locally \emph{coherent} state (with a statistical spread of occupancy $\sqrt{n}$ about $n$) even at intermediate values of $t$.

At intermediate $t$, a superposition of locally correlated states of different $n$ does develop, but due to the linkage between the angular momentum and particle number for the lowest energy FQH states, only two FQH states are superposed in any given superfluid state. The superfluid resembles a hard-core Bose gas of composite particles or holes in the neighboring FQH insulator state, with only a single nonzero component $\psi_m$ of the order parameter, with $m=\Delta L$, the difference in angular momentum of the superposed few-body states. At such a transition with $\epsilon=0$, the bosonic gauge and rotational symmetries are simultaneously spontaneously broken, as the phase of the order parameter determines both a measurable superfluid phase and a rotational orientation to the corresponding few-body superposition of angular momentum eigenstates.

The phase boundary for any $_{n_0}I_\nu$ insulator can be located semi-analytically by expanding the expectation value of the full many-body hamiltonian to second order in the fields $\psi_m$, determining when energy is lowered by nonzero fields. Using a mean-field hamiltonian $\hat{\mathcal{H}}_{MF}=\hat{\mathcal{H}}_{\Omega} + \hat{\mathcal{H}}_{c}-\mu \hat{N}$, where

\begin{eqnarray}
\hat{\mathcal{H}}_{c} = -2t \sum_{n,f,f',m} (\psi_m\tau^{(m)*}_{n,f',f} |n+1,f\rangle\langle n,f'| + \text{c.c.})\nonumber
\end{eqnarray}

\noindent The energy $\langle \hat{\mathcal{H}}\rangle \approx E_0 + \sum_{mm'} \psi_m \mathcal{R}_{mm'} \psi^*_{m'}$, where $E_0$ is independent of the $\psi_m$ and with the matrix $\mathcal{R} = 8N_s t^2 (\gamma-2t\gamma^2)$, where

\begin{equation}
\gamma_{mm'}=\sum_{f} \frac{\tau^{(m')*}_{n_0-1,f,f_0}\tau^{(m)}_{n_0-1,f,f_0}}{\epsilon_{n_0,f_0}-\epsilon_{n_0-1,f}-\mu}+\frac{\tau^{(m')*}_{n_0,f_0,f}\tau^{(m)}_{n_0,f_0,f}}{\epsilon_{n_0,f_0}-\epsilon_{n_0+1,f}+\mu} \nonumber
\end{equation}

\noindent The mean-field energy is lowered when the lowest eigenvalue of $\mathcal{R}$ becomes negative, signifying the phase boundary illustrated by the dotted lines in figure \ref{fig::fig3}a-c. At high $\Omega$ and small but nonzero $t,$  the terms contributing most strongly to $\gamma$ correspond to $\nu=1/2$ (Laughlin) states at particle numbers $n_0\pm 1$; $\mathcal{R}$ has an eigenvalue which first inverts sign for the lower value of $t$ such that $2t\alpha_{n_0-1/2\mp1/2}^2 = 2(n_0-1/2\mp1/2)(1-\Omega/\omega)\mp\mu$ \nobreak

corresponding to the only nonzero tunneling matrix elements between $\nu=1/2$ states $\tau_{n\nu\nu}^{2n}$ (we calculate using DD  $\tau_{n\nu\nu}^{2n}=0.707,0.603,0.546,0.508$, and note that this matrix element asymptotes to $1/2$ as $n$ increases). Thus the boundary from the Laughlin-insulator into the symmetry-broken state has a fixed slope on the $\mu-t$ plane, determined by the structure of the on-site few-body wavefunction, independent of both rotation rate and interaction strength $\eta$, provided $(1-\Omega/\omega) \ll \eta$, and $n$ is sufficiently large.  Analysis of the single-particle Green's function (SOM) shows that the longitudinal masses of particle- and hole-excitations in these insulators are determined by the same parameters as $M_\pm=\hbar^2/t|\alpha_n|^2$, and the energetic gap for their creation at the minimum in their dispersion disappears at these boundaries. These conclusions should persist to larger values of $n$, provided rotation is brought sufficiently close to $\Omega\approx\omega$.

We note that within normalization factors, $\tau^{(m)}_{n,f_1,f_2}$ is equivalent to the matrix element in first quantized form $\int d^{2n} \zeta d^2\zeta\, \Phi^*_{f_1}(\zeta_1 ... \zeta_n)\Phi_{f_2}(\zeta_1 ... \zeta_n| \zeta)\phi_m^*(\zeta)$, where $\Phi_{f_2}(\zeta_1 ... \zeta_n| \zeta)$ represents the $(n+1)$-body state with one particle coordinate symmetrically chosen and set equal to the classical coordinate $\zeta$. Carrying out the sum over $m$ in the chain-coupling Hamiltonian, $\hat{\mathcal{H}}_c= -2t \int d^2\zeta \psi^*(\zeta)\hat{\tau}(\zeta)+c.c.$, where $\psi(\zeta)\equiv\sum_m \psi_m \phi_m(\zeta)$, and $\hat{\tau}(\zeta)=\sum_j \delta(\zeta_j-\zeta)$. Thus the total energy is reduced by creating states of strong overlap between the mean-field and individual particles, tying dynamics of $\psi$ to the transverse motion of ``coordinate-fixing" excitations in the local state. For example, the dynamics and adiabatic manipulation of excitations along the chain in the $_nI_{\nu}$ insulators and $_nS_{\nu}^{\nu}$ superfluids determine the nonzero elements $\tau^{(m)}_{n\nu\nu}$ such that $\langle\hat{\mathcal{H}}_c\rangle\propto -2n Re[ \tilde{t}\langle \int d^2\zeta \prod_j (\zeta_j-\zeta)^{1/\nu}\phi^*_m(\zeta)\psi_m(\zeta)  \rangle]$, the expectation value of a creation operator for $\nu^{-1}$ counter-circulating quasi-holes in the $m=n/\nu$ orbital in the $|n \,\nu\rangle$ state. More generally, the operator $\hat{\tau}$ therefore appears as a type of fusion operator, creating a (fractionalized) hole by statistically distinguishing an individual particle, and in $\hat{\mathcal{H}}_c$, reversibly transferring weight to the coherent field.

Since the energy to promote the mean-field component to a nonzero longitudinal momentum $k$ can be calculated by adiabatically taking $t\rightarrow\tilde{t}\equiv t e^{ika}$, one can see that equivalently the quasihole wavefunction is taken to $\phi_m(\zeta)\propto \zeta^m \rightarrow (\zeta e^{ika/m})^m$, rotating each quasi-hole location through an angle $ka/m=ka\nu/n$. The corresponding quasi-classical orbit of this excitation is therefore screw-like, and includes a component equal to the geometric phase corresponding to a single orbit of a quasi-hole center $\zeta$ around a closed-loop, which defines its fractionalized charge (or mass)\cite{Arovas84}.  The longitudinal current $j=\langle \partial\hat{H}_c(t\rightarrow te^{i\phi})/\partial\phi|_{\phi=0}\rangle$.

In the superfluid phase at high $t$, the lowest energy few-body FQH eigenstates are no longer weighted strongly due to their more even distribution among orbitals $m$, and the superfluid phase reflects a more completely mean-field type state. This is consistent with a picture in which particles must first pay a chain-localization energy $t$ before participating in a strongly correlated on-site state; for modest tunneling strengths the localization energy penalty is sufficiently low to allow atoms to first localize to a lattice site and participate in a strongly correlated state, but at high $t$, this kinetic energy cost is too high, and a delocalized superfluid state is favored.

In the remainder of this letter, we consider novel transport phenomena and properties near criticality through an effective field theory. For finite temperatures, the free energy density $F/N_s=-k_BT\ln Z$ can be constructed using a coherent-state path integral representation of the partition function $ Z=\int  \mathcal{D}[\{b_{im},\psi_{im}\}] e^{-\int_0^\beta \mathcal{L}d\tau}$, with $\beta=1/k_BT$, and Lagrangian

\begin{eqnarray}
\mathcal{L}= \sum_i \mathcal{L}^0_i - \sum_{i,m}( \psi_{im} b^*_{im} + \psi^*_{im} b_{im}) + \sum_{ijm}\frac{\psi_{im}^*\sigma^{-1}_{ij}\psi_{jm}}{t} \nonumber
\end{eqnarray}

\noindent where the complex numbers $b_{im}(\tau)$ label coherent state amplitudes for the $m^{th}$ orbital at site $i$, and $\mathcal{L}^0_i$ is the site-decoupled few-body Lagrangian given explicitly in the supplemental material. We use a generalized Hubbard-Stratonovich transformation \cite{Stratonovich57} to couple the mean-field components $\psi_{im}$, using the tunneling matrix $\sigma_{ij}=\delta_{i,j\pm1}$, and integrate out the $b_{im}(\tau)$ to obtain the partition function $ Z=z_0\int \mathcal{D}[\{\psi_{im}\}] e^{-\int_0^\beta \mathcal{L}_{\psi}d\tau}$. The effective lagrangian can be expanded in powers of slowly varying fields $\psi=(\psi_m(z,\tau))$ and gradients as

\begin{multline}
\mathcal{L}_{\psi} \approx \int dz \, [\psi^* r\psi +  \psi^* \psi^* u \psi \psi \\ + \psi^* k_1 \partial_\tau \psi +\partial_\tau\psi^* k_2\partial_\tau\psi+ \partial_z\psi^* K \partial_z\psi ]\label{eq::Z_psi}
\end{multline}

\noindent where the (matrix) coefficients can be determined from the microscopic parameters by expanding the field-dependent portion of the path-integral, yielding expressions detailed in the SOM dependent on the single- and two-particle Green's functions confined to a single lattice site. The decoupled few-body partition function $z_0=\sum_{nf} \exp{((\mu n-\epsilon_{nf})/k_BT)}$, and one finds generally the matrix $r$ (which forms a complex Hessian for the free energy of the insulator),

\begin{eqnarray}
r_{mm'}&=& \delta_{mm'}/2t - z_0^{-1}\sum_{nff'} \frac{ \tau^{(m)}_{nff'} \tau^{(m')*}_{nff'}}{\epsilon_{nf}-\epsilon_{n+1,f'}+\mu} \\ &\times & (e^{(\epsilon_{n+1,f'}-\mu (n+1))/k_BT}-e^{(\epsilon_{nf}-\mu n)/k_BT})\nonumber
\end{eqnarray}

\noindent Reaching insulator states in the chain requires temperatures and adiabatic manipulation timescales more demanding than does producing FQH states in isolated wells - the finite temperature insulator boundaries are found by setting the lowest eigenvalue of $r$ to zero as illustrated in figure~\ref{fig::fig3}c, displaying a memory of the insulator boundary structure up to temperatures of $0.5\eta\hbar/k_B$ for $\Omega=0.997\omega$, corresponding to $\approx800$pK for physical parameters consistent with the experiment in ref. \cite{Gemelke10}, though we note that this scale increases with stronger interactions and stronger planar confinement.  The parameter $u$ can be extracted from the few-body spectrum as described in the SOM, and directly probes two-particle physics - its explicit calculation permits calculation of critical points in the superfluid phases at higher $t$, as illustrated by the solid lines in figure \ref{fig::fig3}f, which correspond to continuous transitions at saddle points along a new direction of $\psi$ in the free energy introduced by nonzero fields $\psi_{m_0}$.

The remaining coefficients are most transparently recovered from symmetries under two global time-dependent rotation of phases $(b\text{ and }\psi)_{im}\rightarrow (b\text{ and }\psi)_{im}e^{im^s\phi_s(\tau)}$, and corresponding shifts $(\mu,\Omega)_s\rightarrow (\mu,\Omega)_s + i(-1)^s\partial \phi_s / \partial \tau$ for either $s=0$ (bosonic gauge rotation) or $s=1$ (physical rotation about $z$) with $\epsilon=0$. Similar to the single-orbital Bose-Hubbard case $k_\sigma=-\partial^\sigma r/\partial \mu^\sigma$, and for $\epsilon=0$, $k_{1m'm}=(\partial r_{m'm}/\partial \Omega)/m$. The parameter $k_1$ is given by the inverse slope of the phase boundary in the $\mu - t$ plane. Near the centrifugal limit, $k_1$ never vanishes by this calculation, and thus all transitions to the coherent state lie in the universality class of a dilute gas, in contrast to the single-orbital case at low $\Omega$.  More insight into the transition can be found from the single-particle Green's function for the chain using a cumulant expansion\cite{Metzner91} in powers of $t$, similar to the single-orbital case\cite{MPAFisher89,Ohliger12}, but with diagrams carrying orbital indices (this can be performed non-pertubatively in $t$ by summing an infinite set of diagrams\cite{Metzner91,Ohliger12}).

Finally, we note a unique feature of superfluidity in the fractional Hall chain related to interplay of bosonic gauge and rotational symmetries. Using the expansion \ref{eq::Z_psi} above, one can alternately consider the stiffness $\partial^2 F/\partial \phi_s^2$ of the superfluid to ``twists'' of either the phase of the order parameter ($\phi_0$), or a ``torsional'' twist ($\phi_1$) of the gas over its physical length or in time. By standard arguments, the increase in free energy from the former can be used to infer the superfluid density \cite{Fisher73}, while the latter defines a torsional stiffness of the chain. The responses to these two deformations are linked in a manner characteristic of each phase. In the $\nu=1/2$ superfluid at occupation $n_0$, for example, the two twists are physically indistinguishable due to the presence of a single nonzero component $\psi_{m=2n_0}$, with a superfluid phase twist of $\phi_0$ equivalent to a torsional twist of $\phi_1=\phi_0/2n_0$ - with periodic boundary conditions, such as would occur in a chain with its ends joined, this connects a discrete fractional twist of the chain about its axis to its quantized superfluid velocity. This belongs to a family of relations between transport parameters which can be understood by finding the change of free energy due to spatial and temporal gradients of the order parameter phases $\phi_s$, relating the shifts from time-gradients to chemical potential and rotation rate, and spatial gradients to superfluid density and torsional stiffness.

\def\urlprefix{}
\def\url#1{}
\bibliographystyle{apsrev}
\bibliography{FQH_Chain}

\begin{thebibliography}{21}
\expandafter\ifx\csname natexlab\endcsname\relax\def\natexlab#1{#1}\fi
\expandafter\ifx\csname bibnamefont\endcsname\relax
  \def\bibnamefont#1{#1}\fi
\expandafter\ifx\csname bibfnamefont\endcsname\relax
  \def\bibfnamefont#1{#1}\fi
\expandafter\ifx\csname citenamefont\endcsname\relax
  \def\citenamefont#1{#1}\fi
\expandafter\ifx\csname url\endcsname\relax
  \def\url#1{\texttt{#1}}\fi
\expandafter\ifx\csname urlprefix\endcsname\relax\def\urlprefix{URL }\fi
\providecommand{\bibinfo}[2]{#2}
\providecommand{\eprint}[2][]{\url{#2}}

\bibitem[{\citenamefont{Cooper and Wilkin}(1999)}]{Cooper99}
\bibinfo{author}{\bibfnamefont{N.~R.} \bibnamefont{Cooper}} \bibnamefont{and}
  \bibinfo{author}{\bibfnamefont{N.~K.} \bibnamefont{Wilkin}},
  \bibinfo{journal}{Physical Review B (Condensed Matter)}
  \textbf{\bibinfo{volume}{60}}, \bibinfo{pages}{R16279}
  (\bibinfo{year}{1999}).

\bibitem[{\citenamefont{Cooper et~al.}(2001)\citenamefont{Cooper, Wilkin, and
  Gunn}}]{Cooper01}
\bibinfo{author}{\bibfnamefont{N.~R.} \bibnamefont{Cooper}},
  \bibinfo{author}{\bibfnamefont{N.~K.} \bibnamefont{Wilkin}},
  \bibnamefont{and} \bibinfo{author}{\bibfnamefont{J.~M.~F.}
  \bibnamefont{Gunn}}, \bibinfo{journal}{Physical Review Letters}
  \textbf{\bibinfo{volume}{87}}, \bibinfo{pages}{120405}
  (\bibinfo{year}{2001}).

\bibitem[{\citenamefont{Wilkin and Gunn}(2000)}]{Wilkin2000}
\bibinfo{author}{\bibfnamefont{N.~K.} \bibnamefont{Wilkin}} \bibnamefont{and}
  \bibinfo{author}{\bibfnamefont{J.~M.~F.} \bibnamefont{Gunn}},
  \bibinfo{journal}{Physical Review Letters} \textbf{\bibinfo{volume}{84}},
  \bibinfo{pages}{6} (\bibinfo{year}{2000}).

\bibitem[{\citenamefont{Regnault and Jolicoeur}(2003)}]{Regnault03}
\bibinfo{author}{\bibfnamefont{N.}~\bibnamefont{Regnault}} \bibnamefont{and}
  \bibinfo{author}{\bibfnamefont{T.}~\bibnamefont{Jolicoeur}},
  \bibinfo{journal}{Phys. Rev. Lett.} \textbf{\bibinfo{volume}{91}},
  \bibinfo{pages}{030402} (\bibinfo{year}{2003}).

\bibitem[{\citenamefont{Chang et~al.}(2005)\citenamefont{Chang, Regnault,
  Jolicoeur, and Jain}}]{Chang05b}
\bibinfo{author}{\bibfnamefont{C.-C.} \bibnamefont{Chang}},
  \bibinfo{author}{\bibfnamefont{N.}~\bibnamefont{Regnault}},
  \bibinfo{author}{\bibfnamefont{T.}~\bibnamefont{Jolicoeur}},
  \bibnamefont{and} \bibinfo{author}{\bibfnamefont{J.~K.} \bibnamefont{Jain}},
  \bibinfo{journal}{Phys. Rev. A} \textbf{\bibinfo{volume}{72}},
  \bibinfo{pages}{013611} (\bibinfo{year}{2005}).

\bibitem[{\citenamefont{Gemelke et~al.}(2010)\citenamefont{Gemelke, Sarajlic,
  and Chu}}]{Gemelke10}
\bibinfo{author}{\bibfnamefont{N.}~\bibnamefont{Gemelke}},
  \bibinfo{author}{\bibfnamefont{E.}~\bibnamefont{Sarajlic}}, \bibnamefont{and}
  \bibinfo{author}{\bibfnamefont{S.}~\bibnamefont{Chu}} (\bibinfo{year}{2010}),
  \eprint{1007.2677}.

\bibitem[{\citenamefont{Zhang et~al.}(2014)\citenamefont{Zhang, Sreejith,
  Gemelke, and Jain}}]{Zhang14}
\bibinfo{author}{\bibfnamefont{Y.}~\bibnamefont{Zhang}},
  \bibinfo{author}{\bibfnamefont{G.~J.} \bibnamefont{Sreejith}},
  \bibinfo{author}{\bibfnamefont{N.~D.} \bibnamefont{Gemelke}},
  \bibnamefont{and} \bibinfo{author}{\bibfnamefont{J.~K.} \bibnamefont{Jain}},
  \bibinfo{journal}{Phys. Rev. Lett.} \textbf{\bibinfo{volume}{113}},
  \bibinfo{pages}{160404} (\bibinfo{year}{2014}).

\bibitem[{\citenamefont{Paredes et~al.}(2001)\citenamefont{Paredes, Fedichev,
  Cirac, and Zoller}}]{Paredes2001}
\bibinfo{author}{\bibfnamefont{B.}~\bibnamefont{Paredes}},
  \bibinfo{author}{\bibfnamefont{P.}~\bibnamefont{Fedichev}},
  \bibinfo{author}{\bibfnamefont{J.~I.} \bibnamefont{Cirac}}, \bibnamefont{and}
  \bibinfo{author}{\bibfnamefont{P.}~\bibnamefont{Zoller}},
  \bibinfo{journal}{Physical Review Letters} \textbf{\bibinfo{volume}{87}},
  \bibinfo{pages}{010402} (\bibinfo{year}{2001}).

\bibitem[{\citenamefont{Hazzard and Mueller}(2010)}]{Hazzard2010}
\bibinfo{author}{\bibfnamefont{K.~R.~A.} \bibnamefont{Hazzard}}
  \bibnamefont{and} \bibinfo{author}{\bibfnamefont{E.~J.}
  \bibnamefont{Mueller}}, \bibinfo{journal}{Phys. Rev. A}
  \textbf{\bibinfo{volume}{81}}, \bibinfo{pages}{031602}
  (\bibinfo{year}{2010}).

\bibitem[{\citenamefont{Arovas et~al.}(1984)\citenamefont{Arovas, Schrieffer,
  and Wilczek}}]{Arovas84}
\bibinfo{author}{\bibfnamefont{D.}~\bibnamefont{Arovas}},
  \bibinfo{author}{\bibfnamefont{J.~R.} \bibnamefont{Schrieffer}},
  \bibnamefont{and} \bibinfo{author}{\bibfnamefont{F.}~\bibnamefont{Wilczek}},
  \bibinfo{journal}{Phys. Rev. Lett.} \textbf{\bibinfo{volume}{53}},
  \bibinfo{pages}{722} (\bibinfo{year}{1984}).

\bibitem[{\citenamefont{{Stratonovich}}(1957)}]{Stratonovich57}
\bibinfo{author}{\bibfnamefont{R.~L.} \bibnamefont{{Stratonovich}}},
  \bibinfo{journal}{Soviet Physics Doklady} \textbf{\bibinfo{volume}{2}},
  \bibinfo{pages}{416} (\bibinfo{year}{1957}).

\bibitem[{\citenamefont{Metzner}(1991)}]{Metzner91}
\bibinfo{author}{\bibfnamefont{W.}~\bibnamefont{Metzner}},
  \bibinfo{journal}{Phys. Rev. B} \textbf{\bibinfo{volume}{43}},
  \bibinfo{pages}{8549} (\bibinfo{year}{1991}).

\bibitem[{\citenamefont{Fisher et~al.}(1989)\citenamefont{Fisher, Weichman,
  Grinstein, and Fisher}}]{MPAFisher89}
\bibinfo{author}{\bibfnamefont{M.~P.~A.} \bibnamefont{Fisher}},
  \bibinfo{author}{\bibfnamefont{P.~B.} \bibnamefont{Weichman}},
  \bibinfo{author}{\bibfnamefont{G.}~\bibnamefont{Grinstein}},
  \bibnamefont{and} \bibinfo{author}{\bibfnamefont{D.~S.}
  \bibnamefont{Fisher}}, \bibinfo{journal}{Physical Review B (Condensed
  Matter)} \textbf{\bibinfo{volume}{40}}, \bibinfo{pages}{546}
  (\bibinfo{year}{1989}).

\bibitem[{\citenamefont{Ohliger and Pelster}(2013)}]{Ohliger12}
\bibinfo{author}{\bibfnamefont{M.}~\bibnamefont{Ohliger}} \bibnamefont{and}
  \bibinfo{author}{\bibfnamefont{A.}~\bibnamefont{Pelster}},
  \bibinfo{journal}{World Journal of Condensed Matter Physics}
  \textbf{\bibinfo{volume}{3}}, \bibinfo{pages}{125} (\bibinfo{year}{2013}).

\bibitem[{\citenamefont{Fisher et~al.}(1973)\citenamefont{Fisher, Barber, and
  Jasnow}}]{Fisher73}
\bibinfo{author}{\bibfnamefont{M.~E.} \bibnamefont{Fisher}},
  \bibinfo{author}{\bibfnamefont{M.~N.} \bibnamefont{Barber}},
  \bibnamefont{and} \bibinfo{author}{\bibfnamefont{D.}~\bibnamefont{Jasnow}},
  \bibinfo{journal}{Phys. Rev. A} \textbf{\bibinfo{volume}{8}},
  \bibinfo{pages}{1111} (\bibinfo{year}{1973}).

\bibitem[{\citenamefont{Laidlaw and DeWitt}(1971)}]{Laidlaw71}
\bibinfo{author}{\bibfnamefont{M.~G.~G.} \bibnamefont{Laidlaw}}
  \bibnamefont{and} \bibinfo{author}{\bibfnamefont{C.~M.}
  \bibnamefont{DeWitt}}, \bibinfo{journal}{Phys. Rev. D}
  \textbf{\bibinfo{volume}{3}}, \bibinfo{pages}{1375} (\bibinfo{year}{1971}).

\bibitem[{\citenamefont{Wu}(1984)}]{Wu84}
\bibinfo{author}{\bibfnamefont{Y.-S.} \bibnamefont{Wu}},
  \bibinfo{journal}{Phys. Rev. Lett.} \textbf{\bibinfo{volume}{52}},
  \bibinfo{pages}{2103} (\bibinfo{year}{1984}).

\bibitem[{\citenamefont{Zhang}(1992)}]{Zhang91}
\bibinfo{author}{\bibfnamefont{S.~C.} \bibnamefont{Zhang}},
  \bibinfo{journal}{International Journal of Modern Physics B}
  \textbf{\bibinfo{volume}{06}}, \bibinfo{pages}{25} (\bibinfo{year}{1992}).

\bibitem[{\citenamefont{Boninsegni et~al.}(2006)\citenamefont{Boninsegni,
  Prokof'ev, and Svistunov}}]{Svistunov_Prokofiev}
\bibinfo{author}{\bibfnamefont{M.}~\bibnamefont{Boninsegni}},
  \bibinfo{author}{\bibfnamefont{N.~V.} \bibnamefont{Prokof'ev}},
  \bibnamefont{and} \bibinfo{author}{\bibfnamefont{B.~V.}
  \bibnamefont{Svistunov}}, \bibinfo{journal}{Phys. Rev. E}
  \textbf{\bibinfo{volume}{74}}, \bibinfo{pages}{036701}
  (\bibinfo{year}{2006}).

\bibitem[{\citenamefont{Kent and Peifer}(2002)}]{Kent01}
\bibinfo{author}{\bibfnamefont{R.~P.} \bibnamefont{Kent}} \bibnamefont{and}
  \bibinfo{author}{\bibfnamefont{D.}~\bibnamefont{Peifer}},
  \bibinfo{journal}{International Journal of Algebra and Computation}
  \textbf{\bibinfo{volume}{12}}, \bibinfo{pages}{85} (\bibinfo{year}{2002}).

\bibitem[{\citenamefont{Chow}(1948)}]{Chow47}
\bibinfo{author}{\bibfnamefont{W.-L.} \bibnamefont{Chow}},
  \bibinfo{journal}{Annals of Mathematics} \textbf{\bibinfo{volume}{49}},
  \bibinfo{pages}{pp. 654} (\bibinfo{year}{1948}).

\end{thebibliography}


\begin{thebibliography}{0}
\end{thebibliography}

\section*{Notes}

We thank Kaden Hazzard, Tin-Lun Ho, Jainendra Jain, Chaoxing Liu, Marcos Rigol and Yuhe Zhang for helpful conversations. This work was funded by the NSF under award \#1068570.

\section*{Supplementary Material}

\subsection{Verification of Composite Picture for Few-body Laughlin States}

The single-particle states relevant at high rotation rates are taken to be angular momentum eigenstates in the lowest landau level, determined by the single-particle wavefunctions

\begin{equation}
\phi_m(\zeta)=\zeta^m e^{-|\zeta|^2/2}/\sqrt{\pi m!}
\end{equation}

\noindent with $m$ the orbital index and $\zeta=(x+iy)/a_0$ the complex coordinates in two-dimensions, scaled by the transverse harmonic oscillator length. The third dimension is assumed frozen-out, remaining in the lowest motional eigenstate of the potential, which we assume to be harmonically trapped with an oscillator length $a_\Omega$. Contact interactions are included explicitly using the form

\begin{equation}
\hat{V}=\sum_{\{m\}}V_{\{m\}}\hat{a}^\dagger_{i,m_1}\hat{a}^\dagger_{i,m_2}\hat{a}_{i,m_3}\hat{a}_{i,m_4}
\end{equation}

\noindent with

\begin{equation}
V_{\{m\}}=2\pi\delta_{m_1+m_2}^{m_3+m4}(m_1+m_2)!\,2^{m_1+m_2}(\prod_i m_i!)^{-1/2}
\end{equation}

\noindent which describes the scattering of particles between angular momentum orbitals $m_{1...4}$ in the LLL.

The few-body lagrangian for $\epsilon=0$ follows from the hamiltonian as

\begin{eqnarray}
\mathcal{L}^0_i= \sum_{m} \frac{db^*_{im}}{d\tau}b_{im} +((1-\Omega/\omega)m-\mu)|b_{im}|^2  \nonumber \\ + \sum_{\{m\}} \mathcal{V}_{\{m\}} b^*_{im_1}b^*_{im_2}b_{im_3}b_{im_4}
\end{eqnarray}

\subsection{Zero Temperature Phase Boundaries}

The zero-temperature expansion of the ground state energy can be performed by first calculating the perturbed eigen-state

\begin{eqnarray}
\mathcal{N}^{-1}\overline{| n_0, f_0 \rangle}&=&| n_0, f_0 \rangle \nonumber\\
&-& 2t \sum_{fm} |n_0+1,f\rangle \frac{\psi_m\tau^{(m)*}_{n_0,f_0,f} }{\epsilon_{n_0,f_0}-\epsilon_{n_0+1,f}+\mu}\nonumber\\
&-& 2t \sum_{fm} |n_0-1,f\rangle \frac{\psi_m^*\tau^{(m)}_{n_0-1,f,f_0} }{\epsilon_{n_0,f_0}-\epsilon_{n_0-1,f}-\mu}\nonumber\\
\end{eqnarray}

\noindent where $\mathcal{N}$ is a normalization constant. The form of $\mathcal{R}_{mm'}$ then can be found from the expectation value of the full many-body hamiltonian.

\subsection{Calculation of Terms in Landau Expansion of Free Energy}

Terms in the Landau expansion of free energy can be calculated from partial derivatives of the partition function; the derivatives of order $q$ can be related to the $2q$-point thermal Green's functions

\begin{eqnarray}
\prod_{\{m\}}(\frac{\partial}{\partial \psi_{m_1}})...(\frac{\partial}{\partial \psi_{m_q}})(\frac{\partial}{\partial \psi_{m_1}^*})...(\frac{\partial}{\partial \psi_{m_q}^*}) Z\mid_{\{\psi_m\}\rightarrow 0}
\end{eqnarray}

The first two terms in the expansion are found to be

\begin{eqnarray}
Z_0=[\sum_{nf} e^{-(\epsilon_{nf}-\mu n)/k_BT}]^{N_s} = z_0^{N_s}\\
r_{mm'}=(\frac{\delta_{mm'}}{2 t } -  \mathcal{G}^{(1)}_{mm'})
\end{eqnarray}

\noindent where $\mathcal{G}^{(1)}_{mm'}=\langle \int b^*_m(\tau)b_{m'}(\tau') d\tau d\tau' \rangle$ integrates the single-particle thermal Green's function on a lattice site,
which we find from the spectral function

\begin{widetext}
\begin{eqnarray}
\rho_{mm'}(\omega) &=& z_0^{-1}\sum_{nn'ff'}  \langle n f | \hat{a}_m | n' f' \rangle\langle n' f' | \hat{a}^\dagger_{m'} | n f \rangle (e^{(\epsilon_{n'f'}-\mu n')/k_BT}-e^{(\epsilon_{nf}-\mu n)/k_BT}) \delta(\epsilon_{nf}-\epsilon_{n'f'}-\mu(n-n')-\omega)\nonumber \\
&=&  z_0^{-1} \sum_{nff'}  \tau^{(m)}_{nff'} \tau^{(m')*}_{nff'}\delta(\epsilon_{nf}-\epsilon_{n+1,f'}+\mu-\omega) \times (e^{(\epsilon_{n+1,f'}-\mu (n+1))/k_BT}-e^{(\epsilon_{nf}-\mu n)/k_BT})
\end{eqnarray}

\noindent to be

\begin{eqnarray}
\mathcal{G}^{(1)}_{mm'}=z_0^{-1} \sum_{nff'} \frac{ \tau^{(m)}_{nff'} \tau^{(m')*}_{nff'}}{\epsilon_{nf}-\epsilon_{n+1,f'}+\mu}   (e^{(\epsilon_{n+1,f'}-\mu (n+1))/k_BT}-e^{(\epsilon_{nf}-\mu n)/k_BT})\nonumber
\end{eqnarray}

The second-order term in the expansion of the partition function can be expressed as an integral over the two-particle Green's function isolated on a site:

\begin{eqnarray}
u_{\{m\}} &=&\int_0^\beta d^4\,\tau \,G^{(2)}_{\{m\}}(\tau_1\tau_2|\tau_3\tau_4) \\
 &=&  z_0^{-1}\sum_{\{n,f\}} \int_0^{\beta}\!\!\!d\tau_4\int_{\tau_4}^{\beta}\!\!\!d\tau_3\int_{\tau_3}^{\beta}\!\!\!d\tau_2\int_{\tau_2}^{\beta}\!\!\!d\tau_1 \, e^{-\beta h_1} \prod_{j=1}^4  \langle n_j f_j |\hat{\alpha}^{\sigma_j}_{m_j}| n_{(j+1)\text{mod}4} f_{(j+1)\text{mod}4} \rangle e^{(h_j-h_{(j+1)\text{mod}4})\tau_j}
\end{eqnarray}

\noindent where for brevity $h_i=(\epsilon_{n_if_i}-\mu n_i)$, and the $\hat{\alpha}^\sigma=\hat{a}$ or $\hat{a}^\dagger$ according to the $\sigma_i$, which are permuted over pairs of creation and annhilation operators.  To reduce the computational burden, the $n_i,f_i$ states are chosen to match ``rings" of 4-tuples of nonzero elements $\tau^{(m)}_{nf_1f_2}$ and $\tau^{(m)*}_{nf_1f_2}$ found in the few-body spectrum for each set of indices $\{m\}=(m_{1}...m_{4})$.

In order to determine the universality classes of transitions, the expansion must also include spatial and temporal gradients in the $\psi_m$, for which we expand $\psi_m(z,\tau)\approx \psi_m + z\partial \psi_m /\partial z+\tau\partial \psi_m /\partial \tau$. The remaining terms in the Landau expansion of the lagrangian can be found using symmetry under gauge rotation.

\begin{eqnarray}
k_{1mm'}=-\frac{\partial r_{mm'}}{\partial\mu} =z_0^{-1}  \sum_{nff'} \frac{ \tau^{(m)}_{nff'} \tau^{(m')*}_{nff'}}{(\epsilon_{nf}-\epsilon_{n+1,f'}+\mu)^2}  &\times&  (e^{(\epsilon_{n+1,f'}-\mu (n+1))/k_BT}-e^{(\epsilon_{nf}-\mu n)/k_BT}) \\
- \frac{ \tau^{(m)}_{nff'} \tau^{(m')*}_{nff'}}{\epsilon_{nf}-\epsilon_{n+1,f'}+\mu} &\times&  (\frac{-n-1}{k_BT}e^{(\epsilon_{n+1,f'}-\mu (n+1))/k_BT}+\frac{n}{k_BT}e^{(\epsilon_{nf}-\mu n)/k_BT})  \nonumber
\end{eqnarray}

\begin{eqnarray}
k_{2mm'}=-\frac{\partial^2 r_{mm'}}{\partial\mu^2} = z_0^{-1}  \sum_{nff'} \frac{ -2\tau^{(m)}_{nff'} \tau^{(m')*}_{nff'}}{(\epsilon_{nf}-\epsilon_{n+1,f'}+\mu)^3}  &\times&  (e^{(\epsilon_{n+1,f'}-\mu (n+1))/k_BT}-e^{(\epsilon_{nf}-\mu n)/k_BT}) \nonumber\\
+ \frac{ 2\tau^{(m)}_{nff'} \tau^{(m')*}_{nff'}}{(\epsilon_{nf}-\epsilon_{n+1,f'}+\mu)^2}  &\times&  (\frac{-(n+1)}{k_BT}e^{(\epsilon_{n+1,f'}-\mu (n+1))/k_BT}+\frac{n}{k_BT}e^{(\epsilon_{nf}-\mu n)/k_BT}) \nonumber\\
  +\frac{ \tau^{(m)}_{nff'} \tau^{(m')*}_{nff'}}{\epsilon_{nf}-\epsilon_{n+1,f'}+\mu} &\times&  ((\frac{n+1}{k_BT})^2 e^{(\epsilon_{n+1,f'}-\mu (n+1))/k_BT}+(\frac{n}{k_BT})^2 e^{(\epsilon_{nf}-\mu n)/k_BT})  \nonumber
\end{eqnarray}
\end{widetext}

A cumulant expansion can also be used to determine the single particle Green's function - expressing the decoupled (on-site) Green's function at site $i$ diagrammatically as

\begin{centering}
\unitlength = 0.5mm
\begin{fmffile}{output1} 
$G_{1m'm}^0(\omega_s)$=
\parbox{35mm}{\footnotesize
    \begin{fmfchar*}(30,25)
     \fmfleft{i1}
     \fmfright{o1}
     \fmfv{label=$i$,label.angle=90}{v1}
     \fmf{fermion,label=$m,,\omega_s$}{i1,v1}
     \fmf{fermion,label=$m',,\omega_s$}{v1,o1}
     \fmfdot{v1}
     \end{fmfchar*}
}
\end{fmffile}
\end{centering}

\noindent where the indices $m,m'$ represent orbitals, and the matsubara frequency is labeled as $\omega_s$. Explicitly, the on-site Green's function is assembled from the few-body solutions as

\begin{widetext}

\begin{eqnarray}
G_{1m'm}^0(\omega_s)=\sum_{nff'}(\frac{\tau^{m*}_{n-1,ff'}\tau^{m'}_{n-1,ff'}}{\epsilon_{nf}-\epsilon_{n-1,f'}-\mu+i\omega_s}-\frac{\tau^{m*}_{n,ff'}\tau^{m'}_{n,ff'}}{\epsilon_{n+1f'}-\epsilon_{nf}-\mu+i\omega_s})e^{(\mu n -\epsilon_{nf})\beta}
\end{eqnarray}

\noindent One can build a series expansion for the coupled system in powers of $t$ as a sum over connected diagrams of the form

\unitlength = 0.8mm
\begin{fmffile}{output2} 
$G_{1m'm;i',i}(\omega_s)=\sum\limits_{\{j_\ell\},\{ m_\ell\}}$
\parbox{19.2mm}{
    \footnotesize
    \begin{fmfchar*}(24,25)
     \fmfleft{i1}
     \fmfright{o1}
     \fmfv{label=$i$,label.angle=90}{v1}
     \fmf{fermion,label=$m,,\omega_s$}{i1,v1}
     \fmf{fermion,label=$m',,\omega_s$}{v1,o1}
     \fmfdot{v1}
     \end{fmfchar*}
} +
    \parbox{28.8mm}{
    \footnotesize
    \begin{fmfchar*}(36,25)
     \fmfleft{i1}
     \fmfright{o1}
     \fmfv{label=$i$,label.angle=90}{v1}
     \fmfv{label=$i'$,label.angle=90}{v2}
     \fmf{fermion,label=$m,,\omega_s$}{i1,v1}
     \fmf{fermion,label=$m_1,,\omega_s$}{v1,v2}
     \fmf{fermion,label=$m',,\omega_s$}{v2,o1}
     \fmfdot{v1}
     \fmfdot{v2}
     \end{fmfchar*}
 } +
    \parbox{38.4mm}{
    \footnotesize
    \begin{fmfchar*}(48,25)
    \footnotesize
     \fmfleft{i1}
     \fmfright{o1}
     \fmfv{label=$i$,label.angle=90}{v1}
     \fmfv{label=$j_1$,label.angle=90}{v2}
     \fmfv{label=$i'$,label.angle=90}{v3}
     \fmf{fermion,label=$m,,\omega_s$}{i1,v1}
     \fmf{fermion,label=$m_1,,\omega_s$}{v1,v2}
     \fmf{fermion,label=$m_2,,\omega_s$}{v2,v3}
     \fmf{fermion,label=$m',,\omega_s$}{v3,o1}
     \fmfdot{v1}
     \fmfdot{v2}
     \fmfdot{v3}
     \end{fmfchar*}
 } + ...
\end{fmffile}
\end{widetext}

\noindent The summation over internal sites $j_\ell$ is most easily performed in some type of reciprocal space; for reasons which will become clear below, it is illustrative to first perform a $j$-dependent gauge transformation corresponding to a weak twist of frame or bosonic gauge, changing both the underlying basis states $\phi_{jm}\rightarrow\phi_{jm} e^{ijm^s\phi_s}$ and $t\sigma_{jk}\rightarrow t^{jk}_m=t e^{id_{jk}m^s\phi_s} \sigma_{jk}$ with $d_{jk}=\pm\delta_{j,k\pm1}$. One can transform to a fourier-basis $\phi_{k_m m}=\sum_j e^{ik_mjd}\phi_{jm}$, which for the nearest-neighbor coupling yields a dispersion relation $t_{k_m}=t\cos(k_md - m^s\phi_s)$ with $d$ the lattice period, and $k_m$ the momentum in the $m^{th}$ orbital in the limit $\phi_s=0$. Associating a factor of $t_{k_m}$ with each hopping line and performing the sum,

\begin{eqnarray}
G_{1m'm}(k,\omega_s)&=&\nonumber\\ G^0_{1m'm}&+&G^0_{1m' m_1}t_{k_{m_1}} G^0_{1m_1 m}+...\nonumber\\= G^0_1\,\sum_{n=0}^{\infty} g^n &=& G^0_1\, (Q^{-1}\tilde{\Gamma}Q)
\end{eqnarray}

\noindent where $g_{m'm}=t_{k_{m'}}G^0_{1m'm}=(Q^{-1}\Gamma Q)_{m'm}$ is an eigenvalue decomposition with eigenvalues $\Gamma_m$, and the matrix $\tilde{\Gamma}_{m'm}=\delta_{m'm}/(1-\Gamma_m)$.  The representation of $G_1$ can be converted through the expansion $G_1(\chi',\chi)\equiv\mathrm{Tr} (\tilde{\phi}\, G_1)$ using the matrix $\tilde{\phi}_{m'm}=\tilde{\phi}^*_{m'}(\chi')\tilde{\phi}_m(\chi)$ formed from basis states in various representations $\chi$, for example with $\chi=(p,p_z)$ and $\phi_s=0$, one has $\tilde{\phi}_m(p,p_z)=p^{m}\delta(p_z-\hbar k_m)/\pi2^{m/2}$. Permuting the trace,

\begin{eqnarray}
G_1(\chi',\chi)=\mathrm{Tr} [ (Q\, \tilde{\phi} G^0_1 \,Q^{-1})\,\tilde{\Gamma} ]\\
=\frac{1}{t}\mathrm{Tr} [ (q_2^{-1}\phi q_2 q_1^{-1}\Gamma_{0}q_1)\tilde{\Gamma} ]\nonumber
\end{eqnarray}

\noindent where the subscript $0$ indicates setting the $k=(k_m)=0$, the unitary matrices $q_1=Q_0Q^{-1}$ and $q_2=SQ^{-1}$, with $\tilde{\phi}=S^{-1}\phi S$ the eigenvalue decomposition of $\tilde{\phi}$.

In the cylindrically symmetric limit $\epsilon=0$, the matrices $Q,Q_0$ and $q_1$ reduce to the identity, and $q_2=S$, reducing the calculation of the Green's function to

\begin{eqnarray}
G_1(\chi',\chi)&=&\frac{1}{t}\mathrm{Tr} [ \phi ( S \Gamma_{0} \tilde{\Gamma} S^{-1}) ]
\end{eqnarray}

Poles in the Green's function, once analytically continued to real time, describe various excitations. In the decoupled limit, all such excitations have a finite energetic penalty. Once tunnel coupling is introduced, each excitation can delocalize, and the energy as determined by the corresponding pole of $G_1$ becomes dependent on the longitudinal momenta $k=(k_m)$; determining the minimum in energy as a function of $k$ yields the gap, and the lowest order expansion coefficient of $\delta k^2$ away from this point in the pole frequency allows for determination of the mass of the excitation. Presumably at some value of $t$, the mode becomes gapless; the first such mode to become gapless determines the location of the phase boundary, reproducing the condition on the matrix $r$ given in the main text, and defining the lowest energy excitations near the phase boundary.

At low temperature, for $\epsilon=0$, and $\mu$ and $\Omega$ chosen to place the equilibrium thermodynamic ground state in the $_{n_0}I_{\nu_0}$ insulating phase, the dynamical single-particle Green's function for an isolated well reduces to

\begin{eqnarray}
G^0_{1,m'm} \approx -i (\frac{\tau^{m*}_{n_0-1,\nu_0\nu_-}\tau^{m'}_{n_0-1,\nu_0\nu_-}}{\epsilon_{n_0\nu_0}-\epsilon_{n_0-1,\nu_-}-\mu+\omega+i\eta}\nonumber\\-\frac{\tau^{m*}_{n_0,\nu_0\nu_+}\tau^{m'}_{n_0,\nu_0\nu_+}}{\epsilon_{n_0+1\nu_+}-\epsilon_{n_0\nu_0}-\mu+\omega+i\eta})
\label{eq::G10_LT}\end{eqnarray}

\noindent where $\nu_{\pm}$ refer to filling factors of the neighboring particle number states of lowest thermodynamic potential, representing particle- and hole-like excitations in the insulator.  The matrix elements $\tau^{m_-}_{n_0-1,\nu_0\nu_-}=\langle n_0-1, \nu_- |\hat{a}_{m_-}| n_0, \nu_0\rangle\neq0$ only for $m_-=n_0(n_0-1)/2\nu_0-(n_0-1)(n_0-2)/2\nu_-$  and  $\tau^{m_+}_{n_0,\nu_0\nu_+}=\langle n_0, \nu_0 |\hat{a}_{m_+}| n_0+1, \nu_+\rangle\neq0$ only for $m_+=n_0(n_0+1)/2\nu_+-n_0(n_0-1)/2\nu_0$ by conservation of angular momentum.  For the general case $m_+\neq m_-$, and thus the two terms in the sum contribute two relevant eigenvalues $\Gamma_{m_\pm}$, but for the special case $m_+= m_-=m_0$, such as occurs at low rotation rates in all $_{n_0}I_{\infty}$ insulators, only a single eigenvalue $\Gamma_{m_0}$ is relevant.


The Green's function $G_{1m'm}$ in the tunnel-coupled case then exhibits two poles, either from the two terms in $\Gamma_{m_0}$ or each term $\Gamma_{m_\pm}$, defining two low-energy excitations dependent on a single $k_{m_0}$ or two $k_{m_\pm}$ respectively.  Choosing the most compact representation of $G_{1}$ through $\chi=(\ell_z,p_z)$, the eigenvalues $\phi_{m'm}=\delta(p_z-\hbar k_m)\delta(\ell_z-m\hbar)$, and $S$ is the identity.  Performing the sums over values of $m$ and $k_m$, we find the dispersion from the condition $1-\Gamma_m=0$ for cases in which poles reside in different orbitals $m_\pm$ as

\begin{eqnarray}
\omega_\pm&=&\mu\mp(\epsilon_{n_0\pm1,\nu_\pm}-\epsilon_{n_0,\nu_0})\nonumber\\
&\mp&t|\tau^{\ell_z/\hbar}_{n_0-(1\mp1)/2,\nu_0\nu_\pm}|^2\cos{(p_zd/\hbar-(\ell_z/\hbar)^s\phi_s)}\nonumber\\
&\approx & \Delta_\pm + \frac{\hbar^2}{2M_\pm} (p_z-p_{z,0})^2
\end{eqnarray}

\noindent where $\Delta_\pm$ represent the particle/hole gaps at the minima in the dispersion located at $p_{z,0}$, and the $M_\pm$ represent the effective masses.  The gaps vanish along a line in the $\mu-t$ plane governed by

\begin{eqnarray}
\pm\mu-(\epsilon_{n_0\pm1,\nu_\pm}-\epsilon_{n_0,\nu_0})\nonumber\\
=- t|\tau^{\ell_z/\hbar}_{n_0-(1\mp1)/2,\nu_0\nu_\pm}|^2
\end{eqnarray}

\noindent while the masses

\begin{eqnarray}
M_\pm =\hbar^2 / t|\tau^{\ell_z/\hbar}_{n_0-(1\mp1)/2,\nu_0\nu_\pm}|^2
\end{eqnarray}

\noindent show the effect of correlations in slowing the tunneling processes, but never vanish for finite $t$.

The vanishing of a gap $\Delta_\pm$ (in any reference frame determined by the $\phi_s$) is sufficient to predict the location of all phase boundaries separating neighboring $_nI_{1/2}$ insulators from coherent phases, but it is clear from its general form that it cannot predict the location of curved phase boundaries, such as occur for the $_nI_{\infty}$ phases (whose poles inhabit the same orbital $m_0=0$ and thus escape the form above), nor other transitions, such as illustrated in figure \ref{fig::fig3} of the main text for the region between $_4I_{3/4}$ and $_3I_{1/2}$, nor the tip of the $_2I_{1/2}$ phase.

Moreover, in no instance do the gap and the mass vanish at the same point, invariant of Galilean and/or statically twisted reference frames.  In the single-orbital Bose-Hubbard model, such a case does arise at the tri-critical point, and this fact can easily be recovered from the multi-orbital case described here by considering cases where two poles reside in the same orbital $m_0$, wherein the particle and hole excitations hybridize according to equation \ref{eq::G10_LT} to create the modified dispersions

\begin{eqnarray}
\omega^{(c)}_\pm&=&\omega_\pm \pm (\omega_+-\omega_-)[\sqrt{1-h^2}-1]
\end{eqnarray}

\noindent where $\omega_\pm$ are the particle/hole energies found above for the uncoupled case, and

\begin{eqnarray}
h^2=\frac{4t^2|\tau^{\ell_z/\hbar}_{n_0,\nu_0\nu_+}|^2|\tau^{\ell_z/\hbar}_{n_0-1,\nu_0\nu_-}|^2}{(\omega_+-\omega_-)^2}\cos^2(p_zd/\hbar-(\ell_z/\hbar)^s\phi_s) \nonumber
\end{eqnarray}

\noindent

\subsection{Beyond Mean-Local-Field Contributions}

The summations above do not contain all contributing processes to the single-particle Green's function; for example, terms such as

\begin{center}
\unitlength = 0.5mm
\begin{fmffile}{output3} 
\parbox{35mm}{\footnotesize
    \begin{fmfchar*}(50,70)
     \fmfleft{i1}
     \fmfright{o1}
     \fmfv{label=$i$,label.angle=-90}{v1}
     \fmftop{vh}
     \fmf{fermion,left,label=$m,,\omega_s$}{vh,v1}
     \fmf{fermion,left,label=$m,,\omega_s$}{v1,vh}
     \fmf{fermion,label=$m,,\omega_s$}{i1,v1}
     \fmf{fermion,label=$m',,\omega_s$}{v1,o1}
     \fmfdot{v1}
     \fmfdot{vh}
     \end{fmfchar*}
}
\end{fmffile}
\end{center}

\noindent also contribute. In higher dimensions, such diagrams can be argued to contribute negligibly \cite{Ohliger12} to extensive quantities, but in one dimension their exclusion is less trivial, and for the single-orbital Bose Hubbard model strong deviations in the phase boundary shapes are seen between quantum-monte-carlo and mean-field calculations similar to those used above. We leave the evaluation of these terms to future work.

\subsection{Free-Energy of the Insulating States}

To obtain corrections to the ground state energy in the insulator phases due to tunneling, one must calculate the free energy $\mathcal{F}=-(\log{ Z})/\beta$, which requires a sum over all connected vacuum diagrams; to lowest order in $t$, this is due to the single virtual hopping process

\vspace{5mm}
\unitlength = 0.5mm
\begin{fmffile}{output4} 
\begin{equation}
-2\beta\mathcal{F}^{(2)}=\hspace{5mm}
\parbox{12mm}{\footnotesize
    \begin{fmfchar*}(20,25)
     \fmfleft{i1}
     \fmfright{o1}
     \fmfv{label=$j$}{i1}
     \fmfv{label=$k$}{o1}
     \fmf{fermion,left,tension=1.5,label=$m_1,,\tau_1$}{i1,o1}
     \fmf{fermion,left,tension=1.5,label=$m_2,,\tau_2$}{o1,i1}
     \fmfdot{i1}
     \fmfdot{o1}
     \end{fmfchar*}
}\hspace{2mm}=\beta\sum_{jk} \int_0^\beta d\tau \mathrm{Tr} [\tilde{G}^0_{1}(\tau)\tilde{G}^0_{1}(-\tau)]
\end{equation}
\end{fmffile}

\noindent However, since such processes clearly are insensitive to Pierels-phases implementing a spatial twist, we can see that they cannot contribute to any torsional stiffness in the insulating phases, even once taken into the quantum limit. Thus the appearance of torsional stiffness only concurrent with the development of superfluidity is a feature expected to persist beyond the classical mean-field description.

\subsection{Thermodynamic Relations and Transport}

The free energy can be expanded using the fields $\psi_m$ which minimize the action given by the lagrangian in the main text as

\begin{eqnarray}
\beta\Delta F &=&  \sum_{m'ms} \frac{\partial\phi_s}{\partial \tau} (im^s)e^{i(m^s-m'^s)\phi_s}|\psi_{m'}|k_{1m'm}|\psi_{m}| \nonumber\\
&+&  \sum_{m'ms} (\frac{\partial\phi_s}{\partial z})^2 (m'm)^s |\psi_{m'}|K_{m'm}|\psi_{m}| \nonumber\\
&+& \sum_{m'ms} (\frac{\partial\phi_s}{\partial \tau})^2 (m'm)^s |\psi_{m'}|k_{2m'm}|\psi_{m}| \nonumber
\label{eq::ph_exp}\end{eqnarray}

\noindent At the same time, the free energy can be related to changes in chemical potential and rotation rate as

\begin{eqnarray}
\Delta F \approx \sum_s (\partial{F}/\partial{\chi_s})\Delta\chi_s+(\partial^2{F}/\partial{\chi_s}^2)\Delta\chi_s^2/2
\end{eqnarray}

\noindent with $\chi=(\mu,\Omega)$. The changes $\Delta \chi_s$ can be related to temporal gradients of the twist angles $\partial\phi_s/\partial \tau$ using the symmetries under gauge and rotational transforms, and thus each prefactor in the expansion \ref{eq::ph_exp} can be tied to a derivative of the free energy. Since the particle and angular momentum density are $\rho=-\partial{F}/\partial{\mu}$ and $\ell_z=\partial{F}/\partial{\Omega}$ respectively, and the compressibility and moment of inertia $\kappa=\partial^2{F}/\partial{\mu}^2$ and $I=\partial^2{F}/\partial{\Omega}^2$, we have the following relations for physical parameters

\begin{eqnarray}
\rho &=& -\frac{\partial{F_0}}{\partial{\mu}} - \sum_{m'm} \frac{1}{\hbar}|\psi_{m'}| k_{1m'm} |\psi_{m}| \\
\ell &=& \frac{\partial{F_0}}{\partial{\Omega}} + \sum_{m'm} \frac{m}{\hbar}e^{i(m-m')\phi_1} |\psi_{m'}| k_{1m'm} |\psi_{m}| \\
\kappa &=& \sum_{m'm} |\psi_{m'}| k_{2m'm} |\psi_{m}| \\
I &=& \sum_{m'm} |\psi_{m'}|m' k_{2m'm} m|\psi_{m}|
\end{eqnarray}

\noindent For a spatially varying phase with constant gradient, $\phi_s=2\pi z/L$, the change in free energy for $s=0$ defines a superfluid density $\rho_s$ through $\Delta F = \hbar^2\Delta\phi_0^2 \rho_s /2M$, and for $s=1$ defines a torsional stiffness through $\Delta F = S \Delta \phi_1^2 / 2$. One has then also

\begin{eqnarray}
\rho_s &=& \sum_{m'm} \frac{1}{\hbar}|\psi_{m'}| K_{m'm} |\psi_{m}| \\
S &=& \sum_{m'm}  |\psi_{m'}|m' K_{m'm}m |\psi_{m}|
\end{eqnarray}

\subsection{Topological Interpretation}

The time-evolution described by the path-integral in the main text obtained by setting all $\psi_{im}=\psi_m$ can be described by self-consistent evolution of a single two-dimensional system in the presence of a statically defined $\psi$. This leads to a picture of time-evolution in the insulating phases which is (neglecting fluctuations) equivalent to that of an interacting two-dimensional system, in which the domain of the path-integral can be broken into classes of homotopically-equivalent paths.  The same separation of paths can be made in the coherent case. However, since the homotopy groups are a function of the configuration space of the few-body system, we first analyze how this space changes for the cases $\psi=0$ and $\psi\neq0$. We find that while in the insulator phases paths can be described by the standard braid group, the weakly coherent case leads to a entirely different first homotopy group.

The mapping $p_\phi:(\zeta,z)\rightarrow(\zeta e^{i\phi z/L})$ of the chain coordinates ($\zeta,z$) onto to the base space ($\zeta$) is continuous and surjective, such that the chain represents a covering space of the on-site coordinates, and each site a sheet of its base space.  In this case, $p_0$ represents a specific covering map, whose fiber connecting neighborhoods on-site as seen embedded in $\mathbb{R}^3$ is straight. If periodic boundary conditions are imposed on the chain, the twist of any given fiber over the chain becomes discretized by an integer number of revolutions over its length $L$.

Any path in coordinates $\zeta$ can be homotopically lifted to discrete paths in the chain, and the first homotopy (fundamental) group $\pi_1(B)$ describing equivalent paths within the base space $B$ can be related to that of the covering space, the latter a subgroup $p_*(\pi_1(B))$ of the former.  For a single particle, whether in two or higher dimensions $d$, this fact is trivial, as the fundamental group of the simply connected $\mathbb{C}$ or $\mathbb{R}^d$ is simply the identity.

\subsubsection{Path Integral Formulation for Isolated Two-Dimensional Systems and the Insulating Phases}

For a multiparticle system, the decomposition of the mean-field $\psi$ in the coordinates $(\zeta,z)$, however, based on a dynamical mapping from the original symmetrized multiparticle coordinate configuration space $B_n'\equiv(\zeta_{i=1...n})=(\mathbb{C}^n - \mathbb{D})/S_n$, formed by $n$ factors of the complex plane with all points $\mathbb{D}$ removed where any two particle coordinates coincide, properly symmetrized (glued) by forming the quotient with the symmetric group $S_n$\cite{Laidlaw71,Wu84}. Closed paths in this space correspond to trajectories in any few-body path integral similar to that described by $Z$ in the main text, which can be divided into (homotopy) classes of paths which can be smoothly deformed into one another \cite{Laidlaw71}.  The fundamental group $\pi_1(B_n')$ describing equivalent loops in this space is isomorphic to the $n-$particle braid group $\mathcal{B}_n$ generated by the operations $\sigma_i$, clockwise exchanging the particles $i$ and $i+1$. The only relations restricting these operations (the presentation of the braid group) are

\begin{eqnarray}\label{eq::braid_rel}
\sigma_i\sigma_{i+1}\sigma_i=\sigma_{i+1}\sigma_i\sigma_{i+1} \\
\sigma_i\sigma_{j\neq i\pm1}=\sigma_{j\neq i\pm1}\sigma_i \nonumber
\end{eqnarray}

\noindent and the group has the one-dimensional unitary representations $\chi(\sigma_i)=e^{-i\theta}$, with $0\leq\theta<2\pi$, extrapolating continuously from bose ($\theta=0$) to fermi ($\theta=\pi$) statistics. As described in Ref. \cite{Wu84}, any propagator $K$ for a fixed number $n$ particles between initial and final points $q$ and $q'$ in $\mathbb{B}_n'$, can be expressed as a sum over paths in each equivalence class:

\begin{eqnarray}
K(q',\tau'|q,\tau)&=&\\\sum_{\Upsilon\epsilon\pi_1(\mathbb{B}_n')}& \chi(\Upsilon)& \int_{q(\tau)\epsilon\Upsilon} \mathcal{D}[q(\tau)] e^{i\int_\tau^{\tau'} \mathcal{L}[q(\tau)]d\tau} \nonumber
\end{eqnarray}

\noindent Here $\chi(\Upsilon)$ is a phase-factor characterizing each equivalence class $\Upsilon$, which to satisfy composition rules of path integration must form a one dimensional unitary representation of $\pi_1(B_n')$.  Equivalently \cite{Wu84}, the phase factors can be absorbed into the path integral by recognizing that

\begin{eqnarray}
\chi(\Upsilon)=e^{-i\frac{\theta}{\pi}\int d\tau\, \partial_\tau (\sum_{i<j}\angle\zeta_{ij}(\tau))}
\end{eqnarray}

\noindent with $\angle\zeta_{ij}$ the orientation of the relative coordinate between particles $i$ and $j$ in the plane, and the member of the braid group $\Upsilon$ specified by any representative path $q(\tau)=\{\zeta_i(\tau)\}$ in the class. Absorbing these into $K$, they take the form of an additive total time-derivative of the statistical gauge generator $\Lambda\equiv(\theta/\pi)\sum_{i<j} \angle\zeta_{ij}(\tau)$, resulting in the alternate expression for propagators

\begin{eqnarray}
K(q',\tau'|q,\tau)=\int \mathcal{D}[q(\tau)] e^{i\int_\tau^{\tau'} (\mathcal{L}-\partial_\tau\Lambda) d\tau}
\end{eqnarray}

\noindent Following ref. \cite{Wu84}, it is also possible to consider this path integral in the universal covering space $\tilde{\mathbb{B}}_n'$ of $\mathbb{B}_n'$, with final points $q'$ lifted onto separate sheets.  This exploits the isomorphism between the fundamental group $\pi_1(\mathbb{B}_n')$ and the deck transformation group of the mapping from $\tilde{\mathbb{B}}_n'$ to $\mathbb{B}_n'$ to introduce a monodromy action.  In this picture, the propagation of an initial wavefunction $\Phi(q,\tau)$ described by $K$ leads to a multivalued final wavefunction $\Phi(q',\tau')$, with values corresponding to each sheet projected back onto the original configuration space $\mathbb{B}_n'$.

This formulation effectively describes a statistical gauge transformation of the few-body wavefunction $\Phi\rightarrow e^{i\Lambda} \Phi$ leading to the definition of a dynamical gauge field

\begin{eqnarray}
\alpha(\zeta_i)\equiv(\partial_{\text{Re}\zeta_{i}}+i\partial_{\text{Im}\zeta_{i}})\Lambda
\end{eqnarray}

\noindent experienced by each particle arising from its statistical interaction with all others.  In second-quantized form, the gauge field can equivalently be expressed as a solution of the equation

\begin{eqnarray}\label{eq::flux_link}
\partial_{\zeta_r}\alpha_i-\partial_{\zeta_i}\alpha_r = \frac{\theta}{\pi}\rho(\zeta)
\end{eqnarray}

\noindent where the subscripts $r,i$ denote the real and imaginary components, and $\rho$ is the density, expressing the composition of flux and particles. Using the continuity equations for particle currents \cite{Zhang91}, the time derivative $\partial_\tau\alpha=-(\theta/\pi)(j_i+i j_r)$, with $j$ representing particle current.  These equations of motion correspond to an action of the Chern-Simons form

\begin{eqnarray}
\mathcal{L}_\alpha = (\pi/2\theta)(\vec{\alpha} \cdot \nabla \times \vec{\alpha}) - \vec{\alpha}\cdot\vec{j}
\end{eqnarray}

\noindent where $\vec{\alpha}=(\alpha_0,\text{Re} \,\alpha,\text{Im}\, \alpha)$ is a generalization of $\alpha$ above including a time-component $\alpha_0$ as a Lagrange multiplier to enforce condition \ref{eq::flux_link}, and $\vec{j}=(\rho,j_r,j_i)$.  The full system is then described by $\mathcal{L}_0=\mathcal{L}_\alpha + \mathcal{L}_m$, where $\mathcal{L}_m$ can be found from $\mathcal{L}_0$ by the substitution $\vec{A}\rightarrow \vec{A}+\vec{\alpha}$, with $\vec{A}$ the original vector potential,

\begin{eqnarray}
\mathcal{L}_m=\hat{b}^\dagger(\zeta)(i\hbar\partial_\tau-A_0-\alpha_0)\hat{b}(\zeta)\\
-\hat{b}^\dagger(\zeta)\frac{1}{2m}(-i\hbar\partial_\zeta-A-\alpha)^2\hat{b}(\zeta) \nonumber \\
-\hat{b}^\dagger(\zeta)[\frac{m}{2}(\omega^2-\Omega^2)|\zeta|^2-\mu]\hat{b}(\zeta) \nonumber
\end{eqnarray}

Under the choice $\vec{\alpha}=-\vec{A}$, representing a statistical gauge transform under which the new particles experience no net gauge potential, we see that the number of external flux attached per particle is $\nu=b/n=(\pi/\theta)$ - with the condition $\theta=2\pi$, $\nu=1/2$ and $\mathcal{L}_m$ represents a field-free interacting Bose gas.

Before concluding, we note a intuitive picture of the action $\mathcal{S}_\alpha=\int d^2\zeta d\tau\, \mathcal{L}_\alpha$, obtained by gluing the endpoints of the path integral trajectories $q$ and $q'$ to visualize the space as isomorphic to a solid torus.  Integrated over this volume, the action can be expressed as an abelian Chern-Simons three-form $\mathcal{S}_\alpha \propto \int \alpha \wedge d\alpha$ - this form is similar to that discussed in the context of hydrodynamic helicity in three-dimensional fluids \cite{Metzner91} and plasmas, with $\vec{\alpha}$ playing the role of a velocity field in three spatial dimensions.  In these cases, the integration over the three-form forms an integer topological invariant $h=\kappa_i\kappa_j\sum_{ij} \Theta_{ij} + \kappa_j^2\sum_j \Xi_j$ related to the number of links $\Theta_{ij}$ and self-links $\Xi_j$ of individual vortex lines $i,j$ with winding numbers $\kappa_{i,j}$.

\subsubsection{Path Integral Formulation for Weakly-Coupled Two-Dimensional Systems}

Incorporation of the mean-field based on the coupling Hamiltonian $\hat{\mathcal{H}}_c$, and determination of its hydrodynamic properties as described in the main text breaks this process into two topological steps. In the first, described by the elements $\tau(\zeta)$, the $n+1$-particle states are coupled to the $n$-particle states through a (covering) mapping of the underlying manifold from $B_{n+1}'$ to $B_{n}'$, resulting in alternately the matrix elements $\tau^{(m)}$ between states, or their representation as a complex function $\tau(\zeta)$ defined on a simpler manifold $\mathbb{T}$. Once $\tau$ is determined, self-consistent solution with $\psi$ results in a local mean-field model. Finally, lifting this local (two-dimensional) system into the larger space including the chain coordinate $z$ allows for determination of the response of the system to slow spatial and temporal variation of $\psi(\zeta,z)$. Identifying the form of the manifold $\mathbb{T}$ and the function $\tau$ self-consistently with $\psi$ then forms a full solution to the hydrodynamical problem.

The mean field can be incorporated by introducing the field $\psi(\zeta)$ through a modified lagrangian density according to a spatially continuous representation of the same Hubbard-Stravantovich transform used above to obtain the finite temperature phase diagram numerically,

\begin{eqnarray}
\mathcal{L}[\{\psi_i,b_i\}]&=&\sum_i\mathcal{L}_0[b_i]+\sum_i\mathcal{L}_\psi[\psi_i,b_i]+\sum_{ij}\mathcal{L}^{(c)}_\psi[\psi_i,\psi_j]\nonumber
\end{eqnarray}

\noindent with

\begin{eqnarray}
\mathcal{L}_\psi[\psi_i,b_i]&=&-b_i^\dagger(\zeta)\psi_i(\zeta)-b_i(\zeta)\psi_i^*(\zeta)\\
\mathcal{L}^{(c)}_\psi[\psi_i,\psi_j]&=&\psi_i(\zeta)(t\,\sigma_{ij})^{-1}\psi^*_j(\zeta)\nonumber
\end{eqnarray}

\noindent  As above, integrating over all values of $\psi_i(\zeta,\tau)$,

\begin{eqnarray}\label{eq::PI_deloc}
Z=\int \mathcal{D}[\{\psi_i(\zeta,\tau),b_i(\zeta,\tau)\}]\,e^{-\mathcal{S}}
\end{eqnarray}

\noindent with the action

\begin{eqnarray}
\mathcal{S}[\{\psi_i(\zeta,\tau),b_i(\zeta,\tau)\}]=\sum_i\int_0^\beta d\tau d^2\zeta \,\mathcal{L}[\{\psi_i,b_i\}]
\label{eq::action_cont}
\end{eqnarray}

\noindent recovers the full microscopic model and therefore is an exact representation of the original problem in the form of a coherent state path integral with continuous fields.

In order to connect to the picture based on the few-body first-quantized propagator above, and understand how a statistical gauge field or fields may be introduced in the coherent phase, it is helpful to convert the coherent state path integral into a sum over paths in first-quantized form.  To do so, we break the action integral into infinitesimal time steps, replace the coherent fields with $\phi_i(\zeta,\tau)=\langle \{\psi_j(\zeta,\tau),\phi_j(\zeta,\tau)\}| \hat{\phi}_i|\{\psi_j(\zeta,\tau),\phi_j(\zeta,\tau)\} \rangle$, with $|\{\psi_j(\zeta,\tau),\phi_j(\zeta,\tau)\}\rangle$ the coherent state, in which the dynamical auxiliary field $\psi$ has been added to the state label.  Since the path integral \ref{eq::PI_deloc} is formally equivalent to one in which the auxiliary field is represented by an operator $\hat{\psi}_i$ with eigenvalues and states $\hat{\psi}_i|\{\psi_j(\zeta,\tau),\phi_j(\zeta,\tau)\} \rangle = \psi_i(\zeta,\tau)|\{\psi_j(\zeta,\tau),\phi_j(\zeta,\tau)\} \rangle$, we treat both fields identically.

We then expand the path in intermediate states at all times $\tau$ using the identity

\begin{eqnarray}
\mathbb{I}=\sum_n \sum_{(n_b,n_\psi)} \int  &&d^{2n}[\{\zeta,\bar{\zeta}\}]\nonumber\\
 &|&\zeta_1...\zeta_{n_b},\bar{\zeta}_1...\bar{\zeta}_{n_\psi} \rangle\langle\zeta_1...\zeta_{n_b},\bar{\zeta}_1...\bar{\zeta}_{n_\psi}|\nonumber
\end{eqnarray}

\noindent where the second sum is restricted such that $n_b+n_\psi = n$.  The states $|\zeta_1...\zeta_{n_b},\bar{\zeta}_1...\bar{\zeta}_{n_\psi} \rangle$ represent Bose-symmetrized, spatially localized states of definite numbers $n_b$ with coordinates $\zeta_{1...n_b}(\tau)$ and $n_\psi$ with coordinates $\bar{\zeta}_{1...n_\psi}(\tau)$.

The $n_\psi$ quanta are introduced here only as a means for calculation of the path integral above - while they can loosely be interpreted as particles with a longitudinal dependence to their wave function with a distinguishable form to that represented by the $\hat{b}_i$, there is no need to make this association as the problem is fully defined by the choice of action in \ref{eq::action_cont}.  Without further additions to the action, namely topological terms which modify exchange processes, it is clear one must choose the commutators $[\hat{b}_i(\zeta),\hat{b}^\dagger_j(\zeta')]=[\hat{\psi}_i(\zeta),\hat{\psi}^\dagger_j(\zeta')]=\delta_{ij}\delta(\zeta-\zeta')$ and $[\hat{b}_i(\zeta),\hat{\psi}^\dagger_j(\zeta')]=0$.

While the $n_\psi$ quanta are defined here formally through the Hubbard-Stravantovich using the auxiliary coherent states $\psi$, they are reminiscent of the introduction of $n_\psi$ ``worms" in quantum Monte Carlo algorithms \cite{Svistunov_Prokofiev}.  In the latter, the introduction of one or more such entities through ``cuts" of existing or insertions of new world-lines and subsequent movement and joining of their endpoints can connect topologically distinct multi-particle trajectories contributing to the partition function.  Before re-joining endpoints, the trajectories obtained through a single cut/insert and endpoint manipulation represent contributions to the single-particle Green's function, and the inclusion of $n_\psi>1$ such operations represent contributions to higher-order Green's functions.  Given the role of the single-particle Green's function $G_1^{(0)}$ in determining the position of the first phase boundary for fixed chemical potential and increasing tunneling, there is inspiration to consider the classification of its contributing terms according to their topological properties.

To that end, we consider the configuration space of the states $|\zeta_1...\zeta_{n_b},\bar{\zeta}_1...\bar{\zeta}_{n_\psi} \rangle$. Addition of a fixed, transversely localized field $\psi(\zeta)\neq 0$ punctures the complex base space for each quanta of $n_\psi$ at one additional point $\mathbb{P}$, and alters the homotopy classes from $\pi_1(\mathbb{B}_n')$ to $\pi_1(\mathbb{B}_n''[\mathbb{P}_{n_\psi}])$, where

\begin{eqnarray}
\mathbb{B}_n''[\mathbb{P}_s]\equiv\frac{(\mathbb{C}-\mathbb{P}_s)^n - \mathbb{D}}{S_n}
\end{eqnarray}

\noindent are new configuration spaces in which $s$ punctures are placed in the complex-plane for each particle.

For $s=1$, $\pi_1(\mathbb{B}_n''[\mathbb{P}_1])$ is isomorphic to the circular braid group $\mathcal{CB}_n$ of $n$-braidings on an annulus, whose generators $\sigma^c_i$ wind the $i^{th}$ and $(i+1)^{th}$ particles as before, and $\gamma_c$ cyclically permutes the particles around the puncture \cite{Kent01}.  In addition to the relations of Eq.~\ref{eq::braid_rel} applied to the $\sigma^c_i$, for this group

\begin{eqnarray}
\gamma_c^{-1}\sigma^c_i\gamma_c=\sigma^c_{i+1}
\end{eqnarray}

\noindent it is possible to understand this group in a slightly different way - $\mathcal{CB}_n$ has also previously been noted\cite{Chow47,Kent01} to be isomorphic to $\mathcal{D}_{n+1}$, a finite index ($n+1$-coset) subgroup of the $n+1$-particle braid group $\mathcal{B}_{n+1}$, in which a single strand of the original configuration is required to end on its original position \cite{Chow47}. One can thereby alternatively view this as making one particle of the $n+1$-particle path integral statistically distinguishable such that its beginning and end points are fixed - indeed, under the application of the operator $\hat{\tau}(\zeta)$, any state of $n+1$ particles is projected in this way to an $n-$particle state. This reflects the fact that the longitudinal delocalization of a single quanta along the chain effectively identifies a single particle from the few-body system.  From this connection it is clear the homotopic classes of the (weak) mean-field case can correspond to neither braid group $\mathcal{B}_n$ nor $\mathcal{B}_{n+1}$. A unitary one dimensional representation of $\mathcal{CB}_n$ can be found by expanding the mapping $\chi$ to admit $\chi(\gamma_c)=e^{-i\theta_0}$, with $0\leq\theta_0<2\pi$.  This permits tracking windings of particle trajectories about each other and the auxiliary field separately according to any representative path $q_1(\tau)=\{\zeta_1(\tau)...\zeta_{n_b}(\tau),\bar{\zeta}_1(\tau)\}\,\epsilon\,\Upsilon_1$ (similar to the case above) as

\begin{eqnarray}
\chi(\Upsilon_1)=e^{\frac{1}{i\pi}\int d\tau\, \partial_\tau (\theta\sum_{i<j}\angle\zeta_{ij} + \frac{1}{2}\theta_0\sum_{i}\angle(\zeta_{i}-\zeta))}
\end{eqnarray}

\noindent The right-hand side can be seen as a new homotopic invariant corresponding either to self-linkages and cross-linkages between the $\zeta$ and $\bar{\zeta}$ trajectories, or as self-linkages and windings within a thickened toroidal surface.  Setting

\begin{eqnarray}
\Lambda=\frac{\theta}{\pi}\sum_{i<j}\angle\zeta_{ij} + \frac{\theta_0}{2\pi}\sum_{i}\angle(\zeta_{i}-\bar{\zeta_1})
\end{eqnarray}

\noindent and absorbing the phase factors $\chi(\Upsilon)$ into the path integral, a new auxiliary statistical gauge-field emerges as

\begin{eqnarray}
\beta=(\partial_{\bar{\zeta}_{1,r}}+i\partial_{\bar{\zeta}_{1,i}})\Lambda
\end{eqnarray}

\noindent Equivalently, $\alpha$ and $\beta$ can be calculated in second-quantization as

\begin{eqnarray}\label{eq::gauge_src}
\partial_{\zeta_r}\alpha_i-\partial_{\zeta_i}\alpha_r &=& \frac{\theta}{\pi}\rho(\zeta) + \frac{\theta_0}{2\pi}\rho_s(\zeta)\\
\partial_{\zeta_r}\beta_i-\partial_{\zeta_i}\beta_r &=& \frac{\theta_0}{2\pi}\rho(\zeta) \nonumber
\end{eqnarray}

\noindent It will be helpful for reasons which are later apparent to separate $\alpha=\alpha_1+\alpha_2$ into two components

\begin{eqnarray}\label{eq::gauge_src2}
\partial_{\zeta_r}\alpha_{1i}-\partial_{\zeta_{i}}\alpha_{1r} &=& \frac{\theta}{\pi}\rho(\zeta) \\
\partial_{\zeta_r}\alpha_{2i}-\partial_{\zeta_{i}}\alpha_{2r} &=& \frac{\theta_0}{2\pi}\rho_s(\zeta) \nonumber
\end{eqnarray}

\noindent representing the attachment of $\alpha_1$-flux to particles with density $\rho$ and $\alpha_2$-flux to the coherent field $\rho_s=|\psi|^2$, in quanta determined by $\theta$ and $\theta_0$.  Only the $\alpha$ potentials contribute to the effective vector potential seen by particles, while the $\beta$ potential affects the coherent field, and should be included in any ascribed canonical momenta - since it couples to the particle density $\rho$, it represents an interaction between the coherent field and the few-body structure.  Differentiating the above, applying continuity $\partial_\tau \rho=-(\partial_{\zeta_r} j_r +\partial_{\zeta_i} j_i)$,

\begin{eqnarray}\label{eq::mot_gauge}
\partial_\tau\alpha_1 &=& -\frac{\theta}{\pi}(j_i+ij_r) \\
\partial_\tau\beta &=& -\frac{\theta_0}{2\pi}(j_i+ij_r) \nonumber
\end{eqnarray}

\noindent within additive constants.  Equations \ref{eq::gauge_src}-\ref{eq::mot_gauge} can be obtained by minimizing action due to the Lagrangian

\begin{eqnarray}
\mathcal{L}=\mathcal{L}_{\alpha_1}+\mathcal{L}_\beta+\mathcal{L}_m
\end{eqnarray}

\noindent with

\begin{eqnarray}
\mathcal{L}_\beta&=&(\frac{\pi}{\theta_0}) \vec{\beta}\cdot(\nabla\times\vec{\beta}) - \vec{\beta}\cdot\vec{j} \\
\mathcal{L}_{\alpha_1}&=&(\frac{\pi}{2\theta}) \vec{\alpha_1}\cdot(\nabla\times\vec{\alpha_1}) -\vec{\alpha}_1\cdot\vec{j} \nonumber
\end{eqnarray}

\noindent in notation similar to the case for $\alpha$ above. In producing \ref{eq::mot_gauge} from \ref{eq::gauge_src2}, it is necessary to make use of Stoke's theorem, treating \ref{eq::gauge_src2} as the $z$-component of the curl of a three dimensional vector lying in the plane of $\zeta$.  The equivalent expression for $\alpha_2$ is more subtle, if one were to insist here that $\psi$ allow for axial variation $\psi(\zeta,z)$, the time rate of change of coherent density $\partial_\tau \rho_s$ can involve axial currents, and the continuity equation

\begin{eqnarray}
\partial_\tau \rho_s=-\vec{\nabla}\cdot\vec{j_s}
\end{eqnarray}

\noindent must be formed from the divergence of a true three-dimensional current $\vec{j_s}$. Instead, we will make a slowly-varying approximation for the longitudinal coordinate, first treating continuity for $\rho_s$ without the axial component to understand the lagrangian for a homogeneous mean-field case, then proceeding by generalizing that result for axial variation. In this case, the contribution to the Lagrangian due to $\alpha_2$ is

\begin{eqnarray}
\mathcal{L}_{\alpha_2}&=&(\frac{\pi}{\theta_0}) \vec{\alpha_2}\cdot(\nabla\times\vec{\alpha_2}) -\vec{\alpha}_2\cdot\vec{j_s}
\end{eqnarray}

\noindent in similar notation.

The values of $\theta$ and $\theta_0$ have as yet not been chosen, and are in some sense arbitrary choices of statistical gauge.  The ground-state wave-function of the few-body system in the centrifugal limit is simplest under the choice $\theta=2\pi$ when $\rho_s=0$, as this leads to ``condensation" of composites according to the $\frac{1}{2}$-Laughlin form.  For $\rho_s\neq0$, however both $\theta$ and the angle $\theta_0$ must be chosen.  For a single distinguishable trajectory as described here, this is tantamount to finding the ground state and adiabatic dynamics of a single immersed impurity atom with identical characteristics (mass and interaction strengths) to the host atom.  As discussed previously \cite{Zhang14}, near the centrifugal limit this state closely resembles a $1/2-$quasihole, leading to the natural choice of $\theta_0=2\pi$ for those parameters. It is illuminating to consider the effect of adiabatically moving the coordinate $\zeta$ in $\psi(\zeta)$ through a closed loop in the plane.  Under such a process, it is well-known that the full many-body wave-function may accumulate a geometric phase independent of the rate of change of $\zeta$ - in the case of a single quasi-hole in a filling factor $\nu$ state, this corresponds to the Aharanov-Bohm phase accumulated by a single fractionally-charged object in the external magnetic field.  For more general values of $\Omega,\eta$, and $\mu$, the geometric phase may be different or may not exist.  Nevertheless, it is clear that many-body wave-function need not be a single-valued function of $\zeta$, with the domain of $\zeta$ taken to be the complex plane.

\end{document}